\documentstyle[12pt]{article}
\advance\textheight by \topskip
\begin{document}
\vspace{-0.2cm}
\begin{flushright}
MSU--DTP--94/2\\  January 94\\ (Revised: July 94)\\
\vskip2mm
hep-th/9407155
\end{flushright}
\vskip2cm
\begin{center}
{\LARGE \bf EHLERS--HARRISON--TYPE TRANSFORMATIONS\\
\vskip3mm
IN DILATON-AXION GRAVITY}\\
\vskip1cm
{\bf D.V. Gal'tsov} \footnote{ e--mail: dgaltsov@fis.cinvestav.mx,
galtsov@grg.phys.msu.su} {\bf and O.V. Kechkin}
\vskip2mm
{\em Department of Theoretical Physics, Physics Faculty\\
Lomonosov Moscow State University, 119899, Moscow, Russia.}
\end{center}

\vskip 2 cm

\centerline{\bf Abstract}

\begin{quotation}
The ten--parametric internal symmetry group is found in the $D=4$
Einstein--Maxwell--Dilaton--Axion theory restricted to space--times
admitting a Killing vector field. The group includes dilaton--axion
$SL(2,R)$ duality and Harrison--type transformations which
are similar to some target--space duality boosts, but act
on a different set of variables. New symmetry is used to derive a
seven--parametric family of rotating dilaton--axion Taub--NUT dyons.
\vskip5mm
\noindent
PASC number(s): 97.60.Lf, 04.60.+n, 11.17.+y

\end{quotation}

\newpage
\section{Introduction}
Two notable symmetries of the bosonic part of compactified
low--energy heterotic string effective theory were widely discussed recently
\cite{ven}, \cite{sha}, \cite{schw} and used to generate
new classical solutions \cite{sen}, \cite{dow}.
One of them is {\em target space} duality $O(d,d+p)$,
which is valid (in particular) for $D$--dimensional
Einstein--Maxwell--Dilaton--Axion (EMDA) system with $p$ Abelian vector
fields whenever variables are independent of $d$ space--time coordinates
\cite{ven}, \cite{schw}. The (primitive) set of variables
on which the group $O(d,d+p)$ acts consists
of the stringy frame space--time metric, the Kalb--Ramond field
$B_{\mu \nu}$, the vector fields $A_\mu^a, a=1,...,p,$
and the dilaton $\phi$, from which
the corresponding matrix representation is built up. The group mix the metric
with the vector fields, the dilaton, and the axion.
The second symmetry
is {\em dilaton--axion} (or electric--magnetic) duality $SL(2,R)$,
which arises in the case $D=4$ for which
the Kalb--Ramond field can be transformed into
the Peccei--Quinn axion $\kappa$. It says that a pair $\phi, \kappa$
parametrizes the $SL(2,R)/SO(2)$ coset. These two symmetry groups
apparently were regarded as unrelated to each other at least
in the context of the EMDA theory \cite{schw}.

Here we show that for $D=4,\; p=1,\; d=1$, symmetries of both kinds
can be combined
within a larger group. Our approach is similar to that used
earlier for the Einstein--Maxwell (EM) system \cite{nk}, \cite{iw}, \cite{mg}.
It consists in reduction from 4 to 3 dimensions preserving 3--covariance
and involving dualization of non--diagonal metric components
and  magnetic part of the Maxwell tensor.
This leads to the gravity coupled 3--dimensional
{\em sigma--model} (not do be
confused with the initial {\em string} sigma model) with a 6--dimensional
real target space. The latter turns out to possess a 10--parameter
isometry group including the $SL(2,R)$ duality as a subgroup. The group
also contains Harrison--type transfromations, similar to
some target space duality boosts, but now  acting on a
different set of variables related to the primitive variables in a
non point--like way.

Remarkably, our group is larger than the product of both  target space
duality (in this case $O(1,2)$) and  dilaton--axion duality.
Its non--trivial part generalizes Ehlers--Harrison
transfromations known in
the EM theory \cite{eh}, \cite{ha}. The group also contains scale and gauge
transformations. New symmetries open a very simple way
to construct dilaton--axion counterparts to any stationary solution of
the vacuum Einstein equations. As an example we derive a 7--parametric
family of charged rotating Taub--NUT dyons endowed with dilaton and axion
fields. Some future prospects are briefly discussed.
\section{Sigma--model representation}
We start with the $D=4,\;p=1\;$ EMDA action in the Einstein frame
\begin{equation}
S=\frac{1}{16\pi}\int \left\{-R+2\partial_\mu\phi\partial^\mu\phi +
\frac{1}{2} e^{4\phi}
{\partial_\mu}\kappa\partial^\mu\kappa
-e^{-2\phi}F_{\mu\nu}F^{\mu\nu}-\kappa F_{\mu\nu}{\tilde F}^{\mu\nu}\right\}
\sqrt{-g}d^4x,
\end{equation}
where ${\tilde F}^{\mu\nu}=\frac{1}{2}E^{\mu\nu\lambda\tau}F_{\lambda\tau},\;
F=dA\;$, and consider a space--time possessing (at least) one Killing vector
field which we choose here to be time--like.  Then it is standard
to present an interval in terms of a three--metric $h_{ij}$, a rotation
one--form $\omega_i,\; (i,j=1, 2, 3)$ and a scalar $f$ depending
only on space--like coordinates $x^i$,
\begin{equation}
ds^2=g_{\mu\nu}dx^\mu dx^\nu=f(dt-\omega_idx^i)^2-\frac{1}{f}h_{ij}dx^idx^j.
\end{equation}
The vector field may be fully described by 2 real functions: an electric
potential $v$,
\begin{equation}
F_{i0}=\partial_iv/\sqrt{2},
\end{equation}
and a magnetic one $u$,
\begin{equation}
e^{-2\phi}F^{ij}+\kappa {\tilde F}^{ij}=f\epsilon^{ijk}
\partial_ku/\sqrt{2h},
\end{equation}
(note, that $v$ {\em is}, but $u$ {\em is not}
a component of the 4-potential).
Instead of $\omega_k$ a twist potential $\chi$ is then introduced
in accordance with the Einstein constraint equations \cite{iw}
\begin{equation}
\tau_i=\partial_i\chi +v\partial_iu-u\partial_iv,\quad
\tau^i=-f^2\epsilon^{ijk}\partial_j\omega_k/\sqrt{h}.
\end{equation}
Here and below 3--indices are raised and lowered using the metric
$h_{ij}$ and its inverse $h^{ij}$.

New set of variables consists of the 3--metric $h_{ij}$ and
6 ``material'' fields  $\varphi^A=(\kappa, \phi, f , v, u, \chi )$,
$A=1,..., 6$. It is straightforward to check that the field equations
following from the action (1) are identical to the  equations of motion of
a curved space 3--dimensional $\sigma$--model possessing a
6--dimensional target space $\{\varphi^A\}$, together with
the 3--dimensional Einstein equations for $h_{ij}$
with the energy--momentum tensor built from $\varphi^A$.
The corresponding action is
\begin{equation}
S_\sigma=\int \left({\cal R}-{\cal G}_{AB}\partial_i\
varphi^A\partial_j\varphi^B
h^{ij}\right)\sqrt{h}d^3x,
\end{equation}
where ${\cal G}_{AB}$ is the target space metric to be read off from the
line element
\begin{equation}
dl^2 =2d\phi^2+\frac{1}{2}e^{4\phi} d\kappa^2+
\frac{1}{2f^2}\left\{ df^2+(d\chi+vdu-udv)^2\right\}
-\frac{1}{f}\left\{e^{-2\phi}dv^2+e^{2\phi}(du-\kappa dv)^2\right\}.
\end{equation}
Similar representation has been derived for the
stationary EM system by Neugebauer
and Kramer \cite{nk}. Note, that the EMDA theory does not include EM one as a
particular case. Indeed, setting $\phi=\kappa=0$ gives two constraints
$F^2=0=F\tilde F$.
\section{Isometries of the target space }
The target space possess a ten--parametric isometry
group. Seven of its elements can be easily found from the direct inspection
of the metric (7). They include

{\em i) scale transformation}
\begin{equation}
f=f_0e^{2\lambda_1}, \quad \chi =\chi_0e^{2\lambda_1},\quad
v=v_0 e^{\lambda_1}, \quad u=u_0e^{\lambda_1},
\end{equation}
leaving $\kappa $ and  $\phi$ unchanged (here and in what follows $\lambda_s,\,
s=1,..., 10$, are real group parameters);

{\em ii) electromagnetic and gravitational gauge transformations:}
\begin{equation}
u=u_0+\lambda_2,\quad\chi =\chi_0+v_0\lambda_2,
\end{equation}
\begin{equation}
v=v_0+\lambda_3, \quad \chi =\chi_0-u_0\lambda_3,
\end{equation}
\begin{equation}
\chi =\chi_0 +\lambda_4;
\end{equation}
(all other quantities being unchanged) leaving the metric
and the Maxwell tensor invariant;

{\em iii) $SL(2,R)$ dilaton--axion duality subgroup:}
\begin{equation}
z=e^{-2\lambda_5} z_0,\quad
u=u_0e^{-\lambda_5}, \quad v=v_0e^{\lambda_5},
\end{equation}
\begin{equation}
v=v_0,\quad u=u_0+v_0\lambda_6, \quad
z=z_0+\lambda_6,
\end{equation}
\begin{equation}
u=u_0, \quad v=v_0+u_0\lambda_7,\quad z^{-1}=z_0^{-1}+\lambda_7,
\end{equation}
where $z=\kappa +ie^{-2\phi}$ is the complex axidilaton field.

It can be verified that 7 generators of the above transformations form
a closed algebra thus giving no indications on the existence of further
symmetries. However 3 more generators can be found by solving
Killing equations for the target space. They correspond to non--trivial
Ehlers--Harrison--type part of the full symmetry group.

A pair of Harrison--type transformations mix metric functions with
electromagnetic potentials, a dilaton and an axion. Generically,
they produce charged solutions from incharged ones. The first
(electric) leaves invariant the  following quantities
\begin{equation}
f e^{-2\phi}\equiv f_0 e^{-2\phi_0},\quad  \tilde \chi=\chi -
{\tilde u}v\equiv \chi_0 -{\tilde u_0}v_0,
\end{equation}
where $\tilde u =u-\kappa v,$ whereas other variables transform as
\[ {\tilde u}={\tilde u}_0+\tilde\chi_0\lambda_8, \quad
\kappa =\kappa_0+2{\tilde u}_0\lambda_8+\tilde\chi_0\lambda^2_8, \]
\begin{equation}
(\sqrt{f}e^{\phi}\pm v)^{-1}=(\sqrt{f_0}e^{\phi_0}\pm v_0)^{-1}\mp\lambda_8.
\end{equation}

The second (magnetic) also leaves two combinations invariant
\begin{equation}
 q= f^{-1/2} |z| e^\phi\equiv f_0^{-1/2} |z_0| e^{\phi_0},\quad
 p= f^{-1}u^+ u^- \equiv f_0^{-1}u_0^+ u_0^- ,
\end{equation}
where
\begin{equation}
 u^{\pm}=u\pm qf=\frac{u^{\pm}_0}{1-\lambda_9u^{\pm}_0}\, ,
\end{equation}
while other transformations  read
\[  \chi =k_+u^+ + k_- u^-+kqf,\quad v=k_+ \frac{u^+}{u^-} + k_-
\frac{u^-}{u^+},\]
\begin{equation}
z=\frac{fq^2}{dq-i}\,,\quad d=k+k_+ \frac{u^+}{u^-} - k_- \frac{u^-}{u^+}\,,
\end{equation}
\[
 k_{\pm}=\frac{u^{\mp}_0}{2u^{\pm}_0}\left(v_0\pm\frac{\kappa_0f_0e^{2\phi_0}
- \chi_1}{2qf_0}\right), \quad k=\frac{\kappa_0f_0e^{2\phi_0} +
\chi_1}{2qf_0}\,, \]
where $\chi_1=\chi_0-u_0v_0$.

A commutator of two Harrison--type generators gives a generator
of the Ehlers--type transformation. This last transformation, which closes
the full isometry group, has three real
\[ w =e^{2\phi}-v^2 f^{-1}\equiv e^{2\phi_0}-v_0^2 f_0^{-1}, \quad
1-\beta =f^{-1}|\Phi|^2e^{2\phi}\equiv f_0^{-1}|\Phi_0|^2e^{2\phi_0},\]
\begin{equation}
\gamma=f^{-1}(\chi^2+\beta f^2)\equiv f_0^{-1}(\chi_0^2+\beta f_0^2),
\end{equation}
and one complex
\begin{equation}
\nu=v+(if-\chi)\Phi^{-1}\equiv v_0+(if_0-\chi_0)\Phi_0^{-1}
\end{equation}
invariants, where $\Phi=u-zv$, and
\[ f=\chi \xi^{-1}=\gamma(\beta+\xi^2)^{-1}, \quad \xi=\chi_0 f_0^{-1}
-\lambda_{10} \gamma,\]
\begin{equation}
\Phi^{-1}=\Phi_0^{-1}+\nu\lambda_{10},\quad z=z_0-\nu^{-1}(\Phi-\Phi_0).
\end{equation}

There is certain similarity between Harrison--type and some of the string
target space duality transformations. Both generate charged solutions
to the EMDA theory starting from vacuum solutions of the
Einstein equations. However, our group
act on a different set of variables related to the string sigma--model
variables by non point--like transfromations. In the present formulation
dilaton--axion  $SL(2,R)$ duality enters into the same symmetry group.
This group is the symmetry of the sigma--model action (6) and hence that
of the {\em equations of motions} of the initial theory.
Note that in the {\em static} case there seems to exist an analog of Harrison
transformation for the Einstein--Maxwell--Dilaton system (without axion) too
\cite{ho}.
\section{Application to vacuum solutions}
Any solution to the vacuum Einstein equations {\em is} a solution of the
present theory with $v=u=\kappa=\phi=0$. Therefore using the above
transformations
an axion--dilaton counterpart can be found to any stationary vacuum solution.
In this case the above formulas simplify considerably.  The first  Harrison
transformation will read
\begin{equation}
\frac{f}{f_0}=\frac{\chi}{\chi_0}=e^{2\phi}=\frac{1}{1-\lambda_8^2 f_0},\quad
v=\lambda_8 f,\;\; u=\lambda_8 \chi,\;\; \kappa=\lambda ^2 \chi_0.
\end{equation}
If the seed solution is asymptotically flat, and one wishes
to preserve this property, it has to be accompanied by the scale
transformation (8) with the parameter
$e^{2\lambda_1}=1-\lambda^2$. The result can be
concisely expressed in terms of the Ernst potential ${\cal E}=f+i\chi$:
\begin{equation}
{\cal E}=\frac{\sqrt{1-\lambda_8^2}}{\lambda_8}\left(v+iu\right)=
\frac{(1-\lambda_8^2){\cal E}_0}{1-\lambda_8^2 Re {\cal E}_0},\;\;\;
z=i\left(1-\lambda_8^2{\cal E}_0 \right).
\end{equation}

A similar combined transformation via (17)--(19) reads
\begin{equation}
{\cal E}=\frac{\sqrt{1-\lambda_9^2}}{\lambda_9}\left(u-iv\right)=
\frac{(1-\lambda_9^2){\cal E}_0}{1-\lambda_9^2 Re {\cal E}_0},\;\;\;
z=\frac{i}{\left(1-\lambda_9^2{\cal E}_0\right)}.
\end{equation}
In both cases  the metric rotation function is simply rescaled
$\omega_i=(1-\lambda^2)^{-1} \omega_{0i}$,
where $\lambda$ is either  $\lambda_8$,  or $\lambda_9$.

Remarkably, the axidilaton Ehlers--type transformation reduces exactly to
the original Ehlers transformation \cite{eh} when applied to purely vacuum
solutions.
Indeed, in this case $\beta=1$ and from (22) we get ${\cal E}= {\cal E}_0
(1+i\lambda_{10}{\cal E}_0)^{-1}$, while $v, u, \phi, \kappa$ remain zero.

The scale transformation (8) being applied to vacuum solutions reduces to
that of the vacuum Einstein gravity, while the transformations (9)--(14)
trivialize. Therefore, the only non--trivial effect on vacuum solutions
is produced by the Harrison--type transformations. Generically they give
rise to charged configurations endowed with dilaton and axion fields.
\section{Dilaton--axion Kerr--NUT dyon}
Starting with the vacuum Kerr--NUT solution
\begin{equation}
ds_0^2=\frac{\Delta_0-a^2 \sin^2\theta}{\Sigma_0}\left(dt-\omega_0
d\varphi\right)^2-\Sigma_0\left(\frac{dr_0^2}{\Delta_0}+d\theta^2+
\frac{\Delta_0\sin^2\theta}{\Delta_0-a^2\sin^2\theta} d\varphi^2\right),
\end{equation}
where
\[ \Delta_0=r_0(r_0-2M) + a^2 - N_0^2,\quad
\Sigma_0=r_0^2+\delta^2,\quad \delta =a\cos\theta+N_0, \]
\begin{equation}
\omega_0=\frac{2}{a^2\sin^2\theta-\Delta_0}
\left[N_0\Delta_0\cos\theta+
a\sin^2\theta (M_0r_0+N_0^2)\right],
\end{equation}
with the corresponding Ernst potential
${\cal E}_0=1-2(M_0+iN_0)(r_0+i\delta)^{-1}\;$,
($a$ is Kerr rotation parameter, $N_0$ is NUT parameter),
it is a simple matter
to construct its axidilaton counterpart. We will do it
in two steps. First, we perform a constant shift (11)
of the twist potential  ${\cal E}_0 \rightarrow {\cal E}_0 +i\lambda_4$
in order to have one free parameter  more
(this will ensure electric and magnetic charges
in the resulting axidilaton solution to be independent),
and then make either
electric or magnetic Harrison--type transformation accompanied by a suitable
scale transfromation (24), (25) (both lead to the same final form of the
axidilaton
solution). Furthermore, the electromagnetic gauge  freedom (9), (10)
is used to
remove constant asymptotic values of electric and magnetic potentials,
and axidilaton rescaling (12)
is performed to make the dilaton asymptotically zero too.
As  a result, the following electric and magnetic potentials
will be obtained at this step:
\begin{equation}
v^{(0)}=-\frac{2\lambda}{\Sigma}\left(Mr_0+N\delta\right),\;\;
a^{(0)}=\frac{2\lambda}{\Sigma}\left(M\delta -Nr_0-
\frac{\lambda_4 r_-}{2M_0}
\left(Mr_0+N\delta\right)\right),
\end{equation}
where
$M=M_0(1-\lambda^2)^{-1},\; N=N_0(1-\lambda^2)^{-1}\; $ are
rescaled mass and NUT--parameters and
$\Sigma= \Sigma_0 + r_0 r_- -2N_-\delta,\; r_-=2\lambda^2 M, \;
N_-=\lambda^2 N.$

At the second step we consider (26)--(28) as new seed solution
and perform axidilaton duality transformations (12), (14)
with the following parameters
\begin{equation}
e^{-2\lambda_5}=\frac{\left|{\cal M}{\cal Q} \right|^2
e^{-2\phi_{\infty}}}{\left[Im\left({\cal M}z_{\infty}
{\cal Q}\right)\right]^2},
\quad \lambda_7=\frac{Im\left({\cal M}{\cal Q}\right)}
{Im\left({\cal M}z_{\infty}{\cal Q}\right)},
\end{equation}
which are now expressed through the physical quantites: a complex mass
${\cal M}=M+iN$, an electromagnetic charge ${\cal Q}= Q-iP$,
an axidilaton
charge ${\cal D}=D+iA$, and an asymptotic value of the
axidilaton $z_{\infty}$.
Similarly, for $\lambda_4$ one has
\begin{equation}
\frac{\lambda_4 r_-}{2M_0}=
\frac{Re\left({\cal M}z_{\infty}{\cal Q}\right)}
{Im\left({\cal M}z_{\infty}{\cal Q}\right)}.
\end{equation}

The transformed metric can be written in the same form as (26)
\begin{equation}
ds^2=\frac{\Delta-a^2 \sin^2\theta}{\Sigma}
\left(dt-\omega d\varphi\right)^2
-\Sigma\left(\frac{dr^2}{\Delta}+d\theta^2+
\frac{\Delta\sin^2\theta}{\Delta-a^2\sin^2\theta} d\varphi^2\right),
\end{equation}
where now
\[ \Delta=(r-r_-)(r-2M) + a^2 - (N-N_-)^2,\]
\begin{equation}
\Sigma=r(r-r_-)+(a\cos\theta+N)^2 -N_-^2,
\end{equation}
\[
\omega=\frac{2}{a^2\sin^2\theta-\Delta}
\left\{N\Delta\cos\theta+a\sin^2\theta
\left[M(r-r_-)+N(N-N_-)\right]\right\}. \]
The corresponding electric and magnetic potentials and the axidilaton
field are
\begin{equation}
v=\frac{\sqrt{2}e^{\phi_{\infty}}}{\Sigma} Re\left[{\cal Q}
(r-r_-+i\delta)\right],\quad
u=\frac{\sqrt{2}e^{\phi_{\infty}}}{\Sigma} Re\left[{\cal Q}
z_{\infty}(r-r_-+i\delta)\right],
\end{equation}
\begin{equation}
z=\frac{z_{\infty}\rho+{\cal D}z^*_{\infty}}{\rho+{\cal D}},\quad
\rho=r -\frac{{\cal M^*}r_-}{2M}+i\delta.
\end{equation}
Here a new radial coordinate is introduced $r=r_0+r_-$,
parameters $r_-$ and $N_-$ in terms of the physical charges read
\begin{equation}
r_-=\frac{M|{\cal Q}|^2}{|{\cal M}|^2},\quad N_-=\frac{N|{\cal Q}|^2}
{2|{\cal M}|^2},
\end{equation}
and $\delta$ is the same as in (27) (note that $N_0=N-N_-$).
The following expression for the real dilaton function is also useful
\begin{equation}
e^{2\left(\phi-\phi_{\infty}\right)}=\frac{1}{\Sigma}
\left|r+i\delta-\frac{Q{\cal Q}^*}{{\cal M}}\right|^2.
\end{equation}

The solution obtained may be interpreted as the charged rotating
Taub--NUT
dyon in dilaton--axion gravity. It contains seven independent real
parameters:
a mass $M$, a rotation parameter $a$, a NUT--parameter $N$,
electric $Q$ and
magnetic $P$ charges (defined as in \cite{renata} to have
the standard asymptotic normalization of the Coulomb energy),
and asymptotic values of the axion $\kappa_{\infty}$
and the dilaton   $\phi_{\infty}$ (combined in  $z_{\infty}$). The
complex axidilaton charge introduced through an asymptotic expansion
\begin{equation}
z=z_{\infty}-2ie^{-2\phi_{\infty}} \frac{{\cal D}}{r} +
O\left(\frac{1}{r^2}\right),
\end{equation}
is determined by the electromagnetic charge and the complex mass:
\begin{equation}
{\cal D}=-\frac{{\cal Q}^{*2}}{2{\cal M}}.
\end{equation}
Note that this relation is {\em independent} of the rotation parameter $a$.

New family containes as particular cases many previously known
solutions to dilaton--axion gravity. For $N=P=0$
the metric (31), (32) corresponds to
Sen's solution \cite{sen} up to some coordinate transformation
(in this case the axion charge $A=0$). For $a=0$
(31)--(34) coincides (up to a transformation
of the radial coordinate) with the 6--parametric solution
reported recently by Kallosh {\em et al.} \cite{renata}, its
3--parametric subfamily was also found by  Johnson and Myers \cite{jm}.
For $N=0,\; a=0$ we recover the 5--parametric
solution presented by Kallosh and Ortin \cite{rk}, and,
if in addition one of the charges $Q, P$ is zero, the solution
reduces to the Gibbons--Maeda--Garfinkle--Horowitz--Strominger black hole
\cite{gi}. Finally, when $P=Q=0$ we come back to the Kerr--NUT metric (26).

As in vacuum and electrovacuum cases,
for $N\neq 0$ our solution cannot be
properly interpreted as a black hole because of time periodicity
which is to be imposed in presence of the wire singularity \cite{mis}.
We will still conserve the notation $r_H^{\pm}$ for the values
of radial coordinate marking  positions of the surfaces where $\Delta=0$:
\begin{equation}
r_H^{\pm}=M+r_-/2\pm\sqrt{|{\cal M}|^2(1-r_-/2M)^2-a^2}.
\end{equation}
For $N=0$ the upper value $r_H^+$ corresponds to the event horizon of a
black hole. The time--like Killing vector $\partial_t$ becomes null at the
surface $r=r_e(\theta)$,
\begin{equation}
r_e^{\pm}=M+r_-/2\pm\sqrt{|{\cal M}|^2(1-r_-/2M)^2-a^2\cos^2\theta},
\end{equation}
which marks the boundary of a black hole ergosphere in the case  $N=0$.
Inside the 2--surface  $r=r_e(\theta)$ the Killing vector
$\partial_t-\Omega \partial_\varphi$ with some $\Omega=const$ may still
be time--like, the boundary
value of $\Omega$ at $r=r_H^+$ where it becomes null being
\begin{equation}
\Omega_H=\frac{a}{2}\left\{|{\cal M}|^2(1-r_-/2M)+M
\sqrt{|{\cal M}|^2(1-r_-/2M)^2-a^2}\right\}^{-1}.
\end{equation}
For $N=0$ this quantity has a meaning of the angular velocity of the horizon.
The area of the two--surface  $r=r_H^+$ is
\begin{equation}
A=4\pi a/\Omega_H.
\end{equation}

The square root in (40) becomes zero for the family of extremal solutions.
This corresponds to the following relation between parameters
\begin{equation}
|{\cal D}|=|{\cal M}| -a,
\end{equation}
which defines a 4--dimensional hypersurface in the 5--dimensional
space of $Q, P, M, N, a$.
For extremal solutions we have
\[
r_H^{ext}=2M-\frac{aM}{|{\cal M}|},\quad
\Delta^{ext}=\left(r-r_H^{ext}\right)^2,\;\;
\]
\begin{equation}
\omega^{ext}=\frac{2\left\{N \Delta^{ext}\cos\theta + a\sin^2\theta
\left[M\left(r-r_H^{ext}\right)+a|{\cal M}|\right]\right\}}
{a^2\sin^2\theta-\Delta^{ext}}\,,
\end{equation} \[
\Sigma^{ext}=2M\left(r-r_H^{ext}\right)+ \Delta^{ext}-a^2\sin^2\theta
+2a\left(|{\cal M}|+N\cos\theta\right).
\]

The metric for the {\em non--rotating} extremal dilaton--axion
Taub--NUT family reads
\begin{equation}
ds^2=\left(1-2M/r\right) \left(dt+2N\cos\theta d\varphi\right)^2-
\left(1-2M/r\right)^{-1} dr^2 -r(r-2M)\left(d\theta^2+\sin^2\theta
d\varphi^2\right).
\end{equation}
In this case $r_H^{ext}=2M$, this coincides with the curvature singularity.
(Note that it is not so if $a\neq 0$, since then
$\Sigma^{ext}(r_H^{ext})\neq 0$.) For the dilaton we get from (36)
\begin{equation}
e^{2\left(\phi-\phi_{\infty}\right)}=\frac{1}{r(r-2M)}
\left|r-2M+\frac{i(QN-PM){\cal Q}^*}{|{\cal M}|^2}\right|^2.
\end{equation}
Comparing (45) and (46) one can see that generically the string metric
 $ds^2_{string}=e^{2\phi} ds^2$
has non--singular throat structure. However, if
\begin{equation}
QN=PM,
\end{equation}
the dilaton factor (46) has zero as $r\rightarrow 2M$ and the
string metric will have the same structure as in the case
of the static dilaton
electrically charged black hole \cite{gi} (to which our solution
reduces if $N=P=0$). Hence, regular Taub--NUT string throats
form a 3--parametric family corresponding to the hypersurface
$|{\cal D}|=|{\cal M}|$
in the parameter space of $M, N, P, Q$, from which a 2--dimensional subspace
(47) has to be excluded. As it was shown recently by Johnson \cite{j},
some of the family of extremal Taub--NUT solutions
have exact gauged WZW model counterparts.

In the rotating dyon black hole case $N=0,\; a\neq 0$
our solution generalizes (and present in more concise form) the Sen's
solution \cite{sen} to include both electric and magnetic charges and,
consequently, non--zero axion charge. The entropy
can be shown to remain equal to a quarter of the event
horizon area (42),
\begin{equation}
S=\pi\left(\mu +\sqrt{\mu^2-4 J^2}\right),\quad
\mu=2M^2-Q^2-P^2,
\end{equation}
where $J=aM$ is the angular momentum of a hole.
The corresponding temperature is
\begin{equation}
T=\frac{\sqrt{\mu^2-4 J^2}}{4\pi M \left(\mu+ \sqrt{\mu^2-4 J^2}\right)}.
\end{equation}
In the extremality limit $\mu=2|J|$,
the entropy remains finite $S_{ext}=2\pi |J|$ (and vanishing
as $a\rightarrow 0$), while the temperature is zero
in agreement with previous results \cite{hs}.
\section{Conclusion}
Using 3--dimensional sigma--model formulation of the stationary $D=4,\;p=1$
EMDA theory we have found a ten--parametric non--compact internal symmetry
group including dilaton--axion duality and
Ehlers--Harrison--type symmetries.
In a sense, this group provides a unification of target space duality and
dilaton--axion duality in $D=4$.
It also opens a new simple way
to construct stationary solutions to the  $D=4$ EMDA system
by transforming  stationary vacuum solutions as well as already known
solutions to the EMDA theory itself. The above formalism may be
generalized to the case of the space--like Killing vector field too.
Furthermore, it can be shown that the target space (7) is a {\em symmetric
Riemannian space}, on which the isometry group acts transitively.
This reveals a close similarity between the present group
and the Kinnersly $SU(2,1)$
group in the EM theory. It can be anticipated
that in the case of {\em two} commuting Killing vector fields
the system will possess an infinite--dimensional internal symmetry
group analogous to the Geroch--Kinnersley--Chitre
group for electrovacuum. In other words, further
restricted to two dimensions, the EMDA system is likely to become
fully integrable. This will be discussed in a forthcoming paper.
\vskip3mm
\noindent {\bf Acknowledgements} \vskip3mm
This work was supported in parts by the Russian Foundation
for Fundamental Research
grant 93--02--16977, and by the ISF grant M79000.
The authors are also grateful to the Physics
Department of the CINVESTAV del I.P.N., Mexico for hospitality and to
CONACyT for financial support during their visit
while the final version of this paper was written.

\end{document}